\documentclass[fleqn,useAMS,usenatbib,letters]{mn2e}
\usepackage{epsfig,subfigure,amsmath}
\usepackage{color}

\definecolor{purple}{rgb}{1,0,1}

\newcommand{\lcdm}{$\Lambda$CDM}
\newcommand{\hmpc}{$h^{-1}$Mpc}

\newcommand{\beq}{\begin{equation}}
\newcommand{\eeq}{\end{equation}}

\bibpunct[; ]{(}{)}{;}{a}{}{,}

\title[On the observability of cDE with cosmic voids]{On the observability of coupled dark energy with cosmic voids}

\author[P.~M. Sutter et al.]
{
\parbox{\textwidth}{
{P.~M. Sutter}$^{1,2,3}$ \thanks{Email: sutter@iap.fr},
Edoardo Carlesi$^{4}$, 
Benjamin D. Wandelt$^{1,2,5,6}$, and
Alexander Knebe$^{4}$
}
\vspace{0.4cm}\\
\parbox[c]{\textwidth}{
$^{1}$ Sorbonne Universit\'{e}s, UPMC Univ Paris 06, UMR7095, Institut d'Astrophysique de Paris, F-75014, Paris, France \\
$^{2}$ CNRS, UMR7095, Institut d'Astrophysique de Paris, F-75014, Paris, France \\
$^{3}$ Center for Cosmology and Astro-Particle Physics, Ohio State University, Columbus, OH 43210\\
$^{4}$ Departamento de F\'isica Te\'orica, Universidad Aut\'onoma de Madrid, 28049, Cantoblanco, Madrid, Spain \\
$^{5}$ Department of Physics, University of Illinois at Urbana-Champaign, Urbana, IL 61801\\
$^{6}$ Department of Astronomy, University of Illinois at Urbana-Champaign, Urbana, IL 61801\\
}}

\begin{document}

\maketitle

\label{firstpage}

\begin{abstract}
Taking N-body simulations with volumes and particle densities tuned 
to match the SDSS DR7 spectroscopic main sample, 
we assess the ability 
of current void catalogs (e.g., Sutter et al. 2012b) to distinguish 
a model of coupled dark matter-dark energy from \lcdm~cosmology
using properties of cosmic voids.
Identifying voids with the {\tt VIDE} toolkit,
we find no statistically significant differences in the 
ellipticities, but find that coupling produces a population of 
significantly larger voids, possibly explaining the recent result of 
Tavasoli et al. (2013). In addition, we use the universal density profile 
of Hamaus et al. (2014) to quantify the relationship between coupling 
and density profile shape, finding that the coupling produces 
broader, shallower, 
undercompensated profiles for large voids by thinning the walls 
between adjacent medium-scale voids.
We find that these differences are potentially measurable with 
existing void catalogs once effects from survey geometries and 
peculiar velocities are taken into account.
\end{abstract}


\section{Introduction}

Even though a variety of cosmological tests demonstrate that 
the inflation plus cold dark matter (\lcdm) paradigm is 
extremely successful in describing the history and structure 
of the Universe~\citep[e.g.,][]{Reid2012,Planck2013}, 
there are still several features of the large-scale distribution 
of matter that are difficult to explain. One is the so-called 
``void phenomenon'', first noticed by~\citet{Peebles2001}, 
in which cosmic voids --- the deep underdensities in the galaxy 
distribution --- appear emptier than expected from $N$-body simulations.   

The observation of this
 phenomenon motivated the development of models in which 
a dynamical scalar field responsible for dark 
energy~\citep[DE;][]{Peebles:1988, Ratra:1988}
is coupled to the dark matter (DM), giving an additional fifth 
force of nature that would help empty out the voids~\citep{Nusser:2004}.
Other possibilities to explain the void phenomenon have since 
been proposed, including modified 
gravity~\citep[e.g.,][]{Li2009,Clampitt2013,Spolyar2013} 
and an improved understanding of the relationship between 
galaxy formation and environment~\citep{Tinker2009,Kreckel2011}.

Most analyses of coupled DM-DE have focused 
on the 
statistics of overdense regions, such as the halo mass 
function~\citep{Sutter2008},
the galaxy two-point correlation function~\citep{Carlesi:2013a} 
and galaxy cluster gas properties~\citep{Baldi:2010a,Carlesi:2013b}.
However, in these high-density environments it is difficult to distinguish 
effects due to coupling from non-linear evolution and complex baryonic 
physics.

Focusing on underdense regions would appear to be a more natural way to study the void phenomenon. On the theory side, studies have considered
the effect  of coupling in the dark sector on the void number function  ~\citep{Clampitt2013}, density profiles~\citep{Spolyar2013}, 
 and shapes~\citep{Li2009,Li2012}.
Observationally, 
void populations in galaxy surveys can be compared to 
expectations from simulations~\citep[e.g.,][]{Muller2000,Pan2011}.
Most recently,
the study of~\citet{Sutter2013c} found no evidence 
for departures from \lcdm~for a population of voids at higher  redshift ($z\sim 0.4-0.7$), 
but~\citet{Tavasoli2013} noted the existence of 
a large void that appears to be statistically 
incompatible with predictions of \lcdm~$N$-body simulations.	

This comparative inattention to voids themselves can be explained by the 
relative dearth of voids in observations
and the lack of robust void statistical tools that can be used 
to connect theoretical results to observational reality.
However, there have been significant advancements in the past 
few years, including the release of large public void catalogs 
~\citep{Pan2011,Sutter2012a,Sutter2013c} 
from the SDSS galaxy surveys~\citep{SDSS:2009,Ahn2012}.
Secondly, there has been significant efforts to single 
out especially sensitive void properties and make predictions for 
the void signals in 
data~\citep[e.g.,][]{Lavaux2010,Biswas2010,Bos2012,Jennings2013,Sutter2013a}
The combination of enhanced tools and a statistically meaningful 
sample of voids means that predictions of the effects of coupled 
DM-DE within voids can now make direct contact with data.

In this \emph{letter} we provide an initial assessment of the impact 
of coupled DM-DE on void statistics such as number functions, ellipticities, 
and radial density profiles. 
While this work is similar to that of~\citet{Li2011},
we particularly focus on the 
ability of current low-redshift galaxy surveys such as the SDSS 
DR7~\citep{Ahn2012} to distinguish coupled models from \lcdm~ 
with the population of voids identified in their 
limited volumes and galaxy densities~\citep{Pan2011,Sutter2012a}.
We also incorporate the latest theoretical work, such 
as the recently-described universal density profile~\citep[][hereafter HSW]{Hamaus2014}, 
to understand and quantify our results.

In the following section we briefly present the quintessence model, 
its implementation in simulation, and our method for finding voids.
In Section~\ref{sec:effects} we discuss the effects on void 
properties, and conclude in Section~\ref{sec:conclusions} with 
comments on the relevancy for current surveys and outline strategies 
for more complete analyses in the future.

\section{Simulations \& Void Finding}
\label{sec:approach}

Under quintessence the dark energy scalar field $\phi$ has the Lagrangian
\begin{equation}\label{eq:lagrangian}
  L = \int d^4x \sqrt{-g} \left(-\frac{1}{2}\partial_{\mu}\partial^{\mu}\phi 
  + V(\phi)+ m(\phi)\psi_{m}\bar{\psi}_{m} \right),
\end{equation}
where $\phi$ interacts with the matter field $\psi_m$ through the mass term
of the dark matter particles.
In this work we assume the~\citet{Ratra:1988} self interaction potential:
\begin{equation}\label{eq:ratra}
  V(\phi) = V_0\left(\frac{\phi}{M_p}\right)^{-\alpha}
\end{equation}
where $M_p$ is the Planck mass and $V_0$ and $\alpha$
are two parameters that must be fixed by fitting to 
observations~\citep{Wang:2012, Chiba:2013}.

Under this interaction the dark matter particle mass evolves as 
\begin{equation}\label{eq:mass}
  m(\phi) = m_0 \exp{\left(-\beta(\phi)\frac{\phi}{M_p}\right)}.
\end{equation}
This evolution implies that the dark matter particles experience 
an effective gravitational constant of the form~\citep{Baldi:2010a}:
\begin{equation}\label{eq:gravity}
  \tilde{G} = G_{N} (1 + 2\beta^2(\phi))
\end{equation}
where $G_N$ is the standard Newtonian value.
We will fix the interaction term to be constant such that $\beta(\phi)=\beta$.
This leads to a dark matter particle mass that decreases 
as a function of time to its $z=0$ \lcdm~value.
For this work, we contrast a \lcdm~case with a single interacting
model with parameters $V_0=10^{-7}$, $\alpha=0.143$, and 
$\beta=0.099$ (hereafter referred to as cDE).

We used the simulations described in~\citet{Carlesi:2013a} 
and~\citet{Carlesi:2013b} for 
this analysis. 
Briefly, these interactions were implemented with a modified version 
of the Tree-PM code \texttt{GADGET-2}~\citep{Springel:2005}, 
with initial conditions generated using a version of the 
\texttt{N-GenIC} code suitably modified to account for the interactions.
Both simulations had identical initial random phases, 
and were generated using a first order Zel'dovich approximation with 
suitable modifications to account for cDE.
The cosmological parameters used in both \lcdm~and cDE simulations 
were $h=0.7$, $n=0.951$, $\Omega_{dm}=0.224$, $\Omega_b=0.046$, 
and $\sigma_8=0.8$ (normalized at $z=0$) and were constructed to
have dark matter power spectra within current observational 
limits.

These simulations took place in a cubic volume of $250$~\hmpc~per side using 
$1024^3$ DM amd $1024^3$ gas particles. 
We ignored the gas particles and 
randomly subsampled the $z=0.1$ dark matter particles 
to achieve a mean density of 
$\bar n = 4 \times 10^{-3}$ per cubic~\hmpc. This combination of 
simulation volume, density, and redshift approximates the \emph{dim2} 
volume-limited SDSS galaxy sample used in~\citet{Sutter2012a}.  
We subsample the simulations to have identical numbers of particles.
While DM-DE coupling would presumably change the luminosity function of 
galaxies, leading to a change in the total number of 
galaxies in a magnitude-limited survey, we are modeling a 
\emph{volume}-limited survey, which will have identical number counts 
in each scenario
(assuming that the change in galaxy abundances occurs below the magnitude 
threshold of the survey).
Additionally,~\citet{Sutter2013a} 
found that bias does not greatly impact ($< 10\%$) void density profiles 
and abundances, and that the effects of bias are constant across 
difference cosmological models. In summary, to examine the impact of 
DM-DE coupling in a realistic scenario we may ignore galaxy (and halo) bias 
and work only with subsampled dark matter.

We identify voids with the {\tt VIDE} toolkit~\citep{Sutter2014c}, which uses
{\tt ZOBOV}~\citep{Neyrinck2008} 
to construct a Voronoi tessellation of the tracer particles
and apply the watershed transform to group basins into voids.
As in~\citet{Sutter2012a}, we 
remove voids smaller than the mean particle separation 
($6.3$~\hmpc) and 
those with central densities higher than $0.2$ the mean particle 
density $\bar n$. 
Additionally, to limit the growth of voids we set a 
threshold of $0.2 \bar n$ for joining additional zones into voids (see 
~\citet{Neyrinck2008} for a discussion). 
If a void consists of only a single zone then this restriction does not apply.


\section{Results}
\label{sec:effects}

\begin{figure} 
  \centering 
  {\includegraphics[type=png,ext=.png,read=.png,width=\columnwidth]{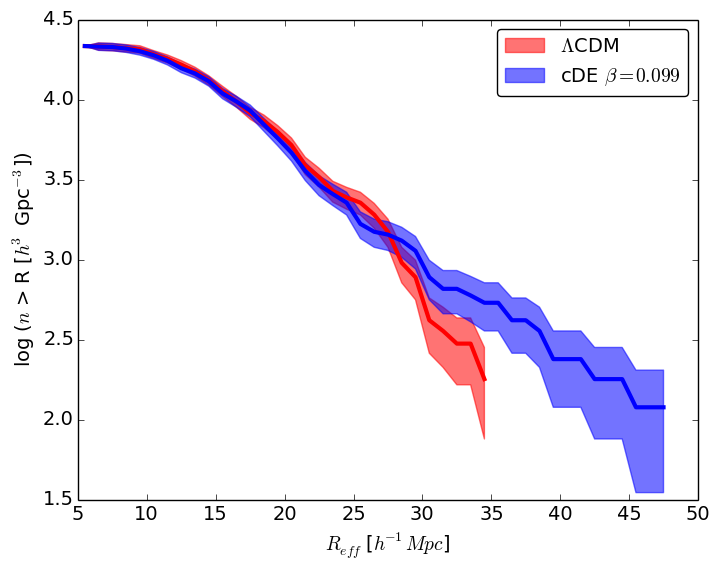}}
  \caption{Cumulative void number functions. 
           Shown are abundances for \lcdm~(red) and cDE (blue) models
           from subsampled $N$-body simulations.
           The solid lines are the measured number functions and the 
           shaded regions are the 1$\sigma$ Poisson uncertainties. 
           For the same $\sigma_8$, cDE results in an excess of large
           scale voids and a deficit of medium scale voids compared
           to \lcdm.
           }
\label{fig:numberfunc}
\end{figure}

Figure~\ref{fig:numberfunc} shows the cumulative number function for 
the \lcdm~and cDE simulations. We immediately note the presence of 
large voids in the cDE simulation, well beyond the largest voids in the 
\lcdm~simulation. However, for smaller void sizes 
($R_{\rm eff} < 20$~\hmpc) the two void 
populations are almost indistinguishable. 
The total number of voids in both models is nearly the same due to our 
fixing of $\sigma_8 = 0.8$, since constraining the freedom of this 
parameter
implies that at least some statistical properties of the cosmic web 
must be retained when departing from \lcdm.
Also, the model we consider here has only modest ($< 2\%$) departures from 
standard gravity.

\begin{figure*} 
  \centering 
  {\includegraphics[type=png,ext=.png,read=.png,width=0.48\textwidth]{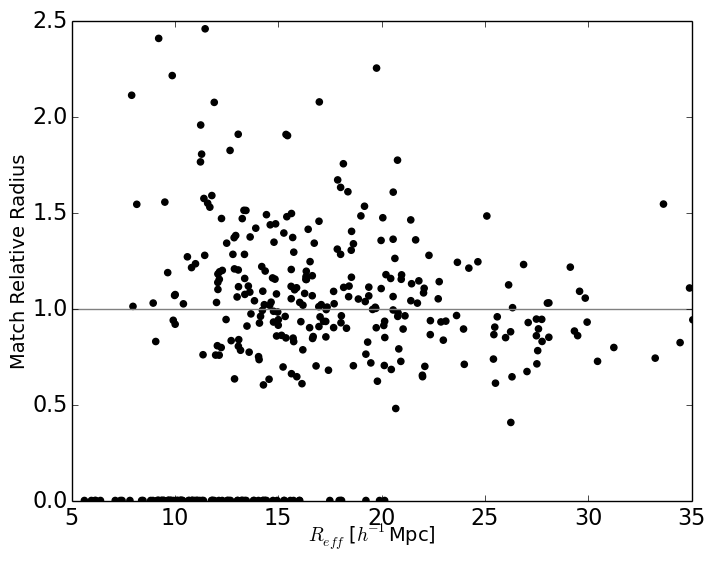}}
  {\includegraphics[type=png,ext=.png,read=.png,width=0.48\textwidth]{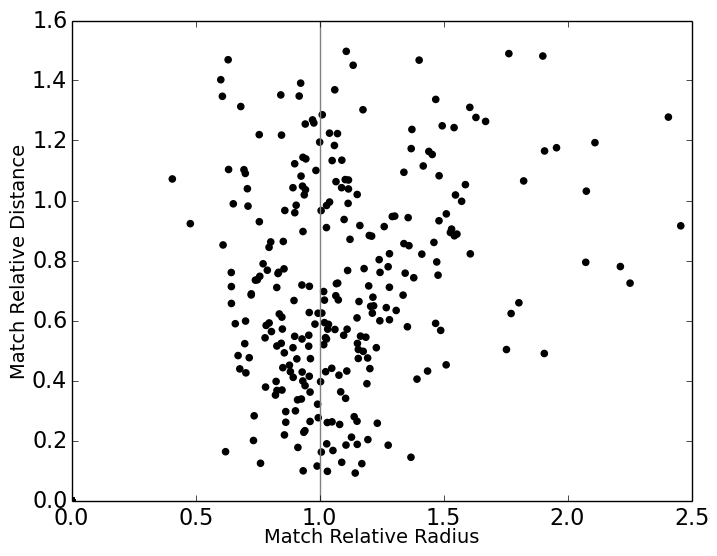}}
  \caption{Relative properties of voids in cDE and \lcdm. Left panel: 
           ratio of radii for voids in cDE matched to voids in \lcdm~as  
           a function of \lcdm~void size. If 
           the void has no match, it is given a relative radius of 0.
           Right panel: Relative distance versus relative radius for 
           matched voids.
           The gray lines marking unity in each panel are to guide the eye.
           Walls between smaller and mid-scale voids are thinned out in 
           the cDE model, making them appear as single, larger voids.
           }
\label{fig:matchprops}
\end{figure*}

To interpret these results we match voids in \lcdm~to voids in the cDE 
simulation using the approach described in~\citet{Sutter2013b}: 
a ``match'' is a corresponding void whose center lies within the 
void under consideration and has the most amount of shared 
particles. The former condition prevents matching to voids in
 the nearby volume that only happen to share a 
few edge particles. Figure~\ref{fig:matchprops} shows the relative radius 
($R_{\rm eff, cDE} / R_{\rm eff, \Lambda CDM}$) and relative macrocenter
distance ($d/R_{\rm eff, \Lambda CDM}$) for the matched voids.
With this insight we see that while the largest voids are largely 
unaffected, small- and medium-scale voids generally 
experience radii inflation of $10-20 \%$, and occasionally dramatic 
increases of up to a factor of two. 
The right panel of Figure~\ref{fig:matchprops} reveals why: 
larger relative radii tend to correspond to large relative distances, 
up to the \lcdm~void effective radius. Thus the walls between 
medium-scale voids are thinned out enough in the presence of cDE 
to allow the watershed to merge them together into a single 
larger void.
This explains the feature at 
$\sim 25$ \hmpc~in the cDE number function: these \lcdm~voids are 
merging together to form the largest cDE voids.
The primary cause of this thinning out, whether the modified expansion 
history or the fifth force itself, requires more investigation.

\begin{figure*} 
  \centering 
  {\includegraphics[type=png,ext=.png,read=.png,width=0.32\textwidth]{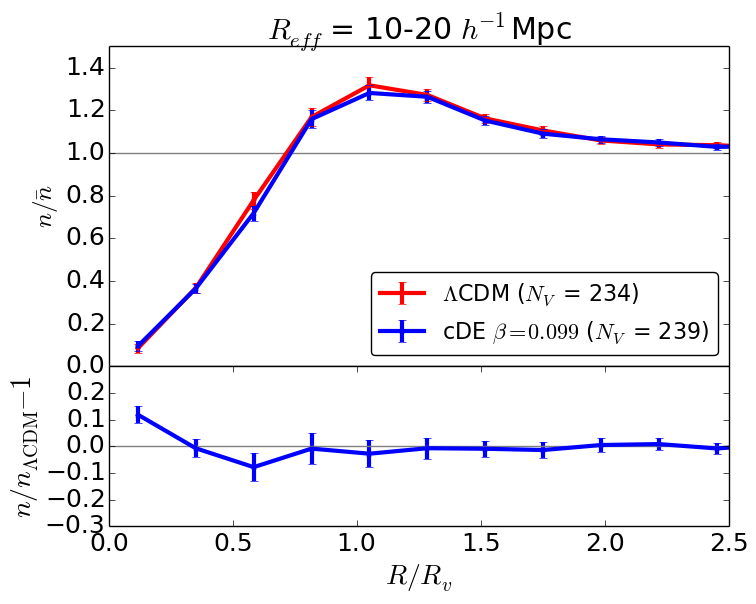}}
  {\includegraphics[type=png,ext=.png,read=.png,width=0.32\textwidth]{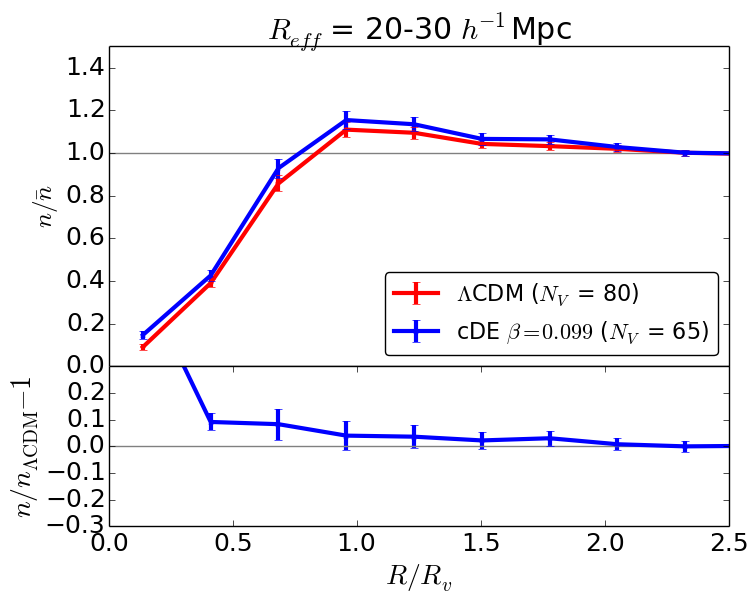}}
  {\includegraphics[type=png,ext=.png,read=.png,width=0.32\textwidth]{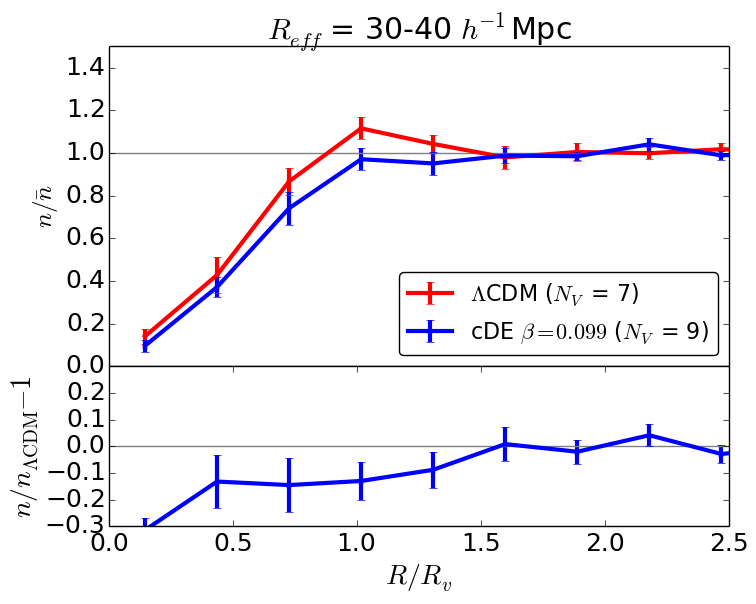}}
  \caption{One-dimensional radial density profiles of stacked voids with 
           1$\sigma$ bootstrapped uncertainties
           (points with error bars). 
           The bottom of each panel shows the cDE profile relative to the 
           \lcdm~one. The solid lines are to guide the eye.
           Shown are profiles from the \lcdm~(red) and cDE (blue) 
           simulations. 
           The legend indicates the number of voids $N_v$ in each stack.
           Coupling enhances the compensation walls of medium-scale 
           voids but diminishes the walls of larger voids.
           }
\label{fig:1d_profile}
\end{figure*}

In Figure~\ref{fig:1d_profile} we show one-dimensional radial profiles
for all samples in radius bins of width 10 \hmpc.
To compute the profiles we take all voids in the radius bin, align all their 
macrocenters,
and measure the total density in thin spherical shells. 
We normalize each
density profile to the mean number density of the sample and show
all profiles as a function of the relative radius, $R/R_v$, where $R_v$
is the median void size in the stack. We do not individually rescale 
the voids since that tends to dampen the compensation 
region~\citep{Sutter2012b}, 
which we wish to highlight in this analysis. 

The profiles in each stack follow the same overall structure 
(a deeply-underdense core, a steep wall, an overdense ``compensation'' 
shell, and a flattening to the mean density); however, there are some 
contrasts between \lcdm~and cDE voids. 
First, while there are almost no differences between \lcdm~and cDE voids 
at the smallest scales,
greater
discrepancies appear for the larger ($> 20$ \hmpc) stacks.
From 20 to 30 \hmpc, cDE voids have higher 
compensation shells, but after 30 \hmpc~the cDE voids are clearly 
larger and flatter (i.e., lower density contrast between wall 
and center). This difference in the largest stack is statistically 
highly significant.

\begin{figure} 
  \centering 
  {\includegraphics[type=png,ext=.png,read=.png,width=0.48\textwidth]{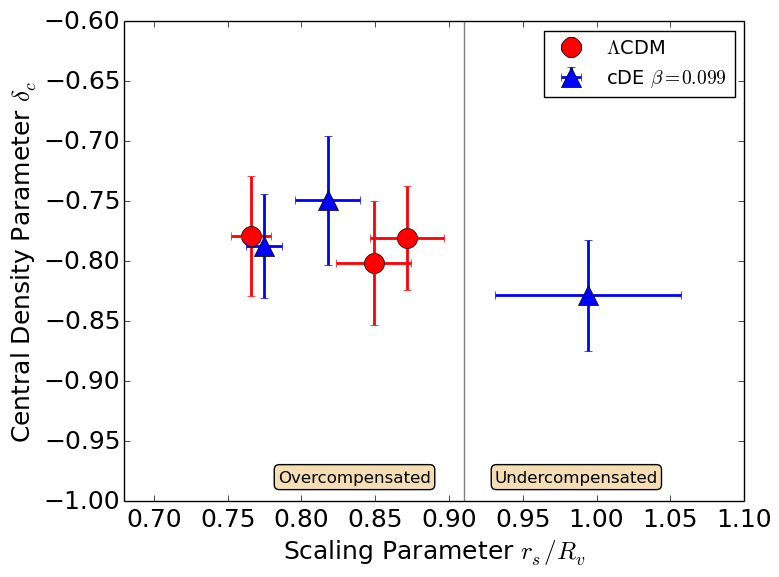}}
  \caption{
           Best-fit values and $1 \sigma$ uncertainties
           for all void stacks studied using the profile given 
           by Eq. (\ref{eq:profile}). 
           Shown are values from \lcdm~(red circles) and
           cDE (blue triangles).
           The thin grey line depicts
           compensation scale.
           For each sample from left to right the points are for the 
           $10-20$, $20-30$, and $30-40$~\hmpc~stacks.
           }
\label{fig:profile_fits}
\end{figure}

To quantify and understand these differences, we fit all the profiles 
to the universal function presented in HSW:
\begin{equation}
\frac{n}{\bar{n}} (r) = \delta_c \frac{1- (r/r_s)^{\alpha(r_s)}}
                             {1 + (r/R_v)^{\beta(r_s)}} + 1,
  \label{eq:profile}
\end{equation}
While there are four parameters total in the model, HSW
describe a two-parameter reduced model where
\begin{eqnarray}
  \alpha(r_s) & \simeq & -2.0 (r_s/R_v) + 4.0 \\
\beta(r_s) & \simeq & \left\{ \begin{array}{rl}
17.5 (r_s/R_v) - 6.5 &\mbox{ if $r_s/R_v < 0.91$} \\
-9.8 (r_s/R_v) + 18.4 &\mbox{ if $r_s/R_v > 0.91$}.
\end{array} \right.
  \label{eq:parmfits}
\end{eqnarray}
This two-parameter model describes all but the largest voids very 
accurately, and is appropriate for the analysis here~\citep{Sutter2013a}.
There are two free parameters to this model: $r_s$, the radius
at which the profile reaches mean density, and $\delta_c$,
the density in the central core. Figure~\ref{fig:profile_fits} shows
all best-fit values of $\delta_c$ and $r_s$ for all stacks in both 
simulations.

The fitting parameters elucidate the relationship between DM-DE 
coupling and profile shape. 
All voids in all models maintain roughly the same central 
density of $\delta_c \sim -0.8$, regardless of size.
However, the evolution of the scaling radius as a function of void 
size is significantly different between the two models.
Indeed, the largest cDE voids in this volume become undercompensated, whereas 
no \lcdm~voids reach the necessary scales. 
Since almost all \lcdm~voids can be matched to a cDE counterpart, 
Figure~\ref{fig:matchprops} also allows us to interpret these results:
the larger cDE voids are actually merged --- not enlarged --- \lcdm~voids, 
making the voids appear uniformly larger without greatly 
affecting the central densities. As in the profiles themselves, 
the differences between individual profiles are only statistically 
significant for the largest radius bin.
 However, there are relatively
few voids here, and we may be underestimating the true uncertainty.

We also examined ellipticities using the inertia tensor method 
as described in~\citet{Bos2012} and~\citet{Sutter2013a}. However, 
as~\citet{Bos2012} discovered, in sparse populations such as galaxies 
it is very difficult to statistically separate \lcdm~from 
alternative cosmologies using void shapes. We found no significant
distinctions in the ellipticities.

\section{Conclusions}
\label{sec:conclusions}

We have examined the effects on voids due to coupling between dark matter 
and dark energy with realistic galaxy survey volumes and tracer densities
and provided an initial assessment of the feasibility of current
surveys to detect the coupling with voids.
We have found that the coupling produces much larger 
voids compared to \lcdm~mostly by merging
medium-scale voids.
Additionally, we have quantified the effects of coupling on the 
radial density profiles by finding the best fits to the 
analytic HSW profile, and found that DM-DE coupling 
can more easily make voids underdense.

Traditional probes of large-scale structure such as the 
power spectrum have difficulty differentiating cDE models 
from \lcdm~\citep[e.g.,][]{Carlesi:2013a}, but voids are exceptionally 
powerful discriminating tools. 
We have studied a relatively weak coupling strength; thus even a null 
result would be informative about the capabilities of void properties to
distinguish these models. 
However, even with limited survey volumes and only $\sim 400$ total voids 
the number functions are distinguishable 
in a statistically significant manner.
Density profiles of the largest voids, despite the relatively few 
number of voids, also provide measurable differences in the 
scaling radius $r_s$ of the HSW profile, although future simulations 
with larger volumes will be needed to verify the precise 
statistical significance of these profile differences.

The void population we have studied is fairly representative of --- and 
accessible with --- current low-redshift galaxy
surveys~\citep{Sutter2012a}. 
These results may explain the large void identified by~\citet{Tavasoli2013}. 
In addition, using 
Halo Occupation Distribution modeling~\citep{Berlind2002} and 
accounting for survey geometries,~\citet{Sutter2013c} was able 
to match \lcdm~simulations to observed void populations. 
Thus, coupled DM-DE may already be measurable with 
current data sets.
However, for a complete comparison 
of the void abundance we must include mask effects~\citep{Sutter2013c}.
We also have not included the effects of galaxy bias and distortions to 
the density profiles from peculiar velocities. However, we can 
use techniques such as those presented by~\citet{Pisani2013} 
to construct the real-space profile without modeling.
We will save a more detailed comparison and measurement of a 
constraint for future work.

This study is only an initial assessment comparing one cDE model to \lcdm, 
using simulations optimized to study the properties of 
high-density clusters. 
We also examined other coupling strengths (Eq.~\ref{eq:gravity}) but did not 
find significant differences among the models with this limited void 
population. We are preparing larger simulations that 
will allow us to examine the 
detailed relationship between coupling strength and void properties
and assess the ability of high-redshift galaxy surveys 
such as BOSS~\citep{Dawson2013} to probe these cosmologies using voids.
In addition, future galaxy surveys will only serve to increase 
the statistical significance of these differences, leading to 
ever-further constraints on these models.

\section*{Acknowledgments}

PMS and EC thank Alexander Kusenko and the NSF for the PACIFIC2013 workshop,
where this work was initiated. 
The authors thank the referee for the highly constructive 
comments that substantially improved this letter, as well as 
Nico Hamaus, Alice Pisani, and Ravi Sheth.
BDW 
acknowledges funding from an ANR Chaire d'Excellence (ANR-10-CEXC-004-01),
the UPMC Chaire Internationale in Theoretical Cosmology, and NSF grants AST-0908
902 and AST-0708849.
This work made in the ILP LABEX (under reference ANR-10-LABX-63) was supported by French state funds managed by the ANR within the Investissements d'Avenir programme under reference ANR-11-IDEX-0004-02. 

AK is supported by the {\it Ministerio de Econom\'ia y Competitividad} (MINECO) in Spain through grant AYA2012-31101 as well as the Consolider-Ingenio 2010 Programme of the {\it Spanish Ministerio de Ciencia e Innovaci\'on} (MICINN) under grant MultiDark CSD2009-00064. He also acknowledges support from the {\it Australian Research Council} (ARC) grants DP130100117 and DP140100198. He further thanks Ghosts I've Met for winter's ruin.

\footnotesize{
  \bibliographystyle{mn2e}
  \bibliography{voidcde}

\begin{thebibliography}{}

\bibitem[\protect\citeauthoryear{{Abazajian} et~al.,}{{Abazajian}
  et~al.}{2009}]{SDSS:2009}
{Abazajian} K.~N.,  et~al., 2009, \apjs, 182, 543

\bibitem[\protect\citeauthoryear{{Ahn} et~al.,}{{Ahn}  et~al.}{2012}]{Ahn2012}
{Ahn} C.~P.,  et~al., 2012, \apjs, 203, 21

\bibitem[\protect\citeauthoryear{{Baldi}, {Pettorino}, {Robbers} \&
  {Springel}}{{Baldi} et~al.}{2010}]{Baldi:2010a}
{Baldi} M.,  {Pettorino} V.,  {Robbers} G.,    {Springel} V.,  2010, \mnras,
  403, 1684

\bibitem[\protect\citeauthoryear{{Berlind} \& {Weinberg}}{{Berlind} \&
  {Weinberg}}{2002}]{Berlind2002}
{Berlind} A.~A.,  {Weinberg} D.~H.,  2002, \apj, 575, 587

\bibitem[\protect\citeauthoryear{Biswas, Alizadeh \& Wandelt}{Biswas
  et~al.}{2010}]{Biswas2010}
Biswas R.,  Alizadeh E.,    Wandelt B.,  2010, \prd, 82

\bibitem[\protect\citeauthoryear{{Bos}, {van de Weygaert}, {Dolag} \&
  {Pettorino}}{{Bos} et~al.}{2012}]{Bos2012}
{Bos} E.~G.~P.,  {van de Weygaert} R.,  {Dolag} K.,    {Pettorino} V.,  2012,
  ArXiv e-prints: 1205.4238

\bibitem[\protect\citeauthoryear{{Carlesi}, {Knebe}, {Lewis}, {Wales} \&
  {Yepes}}{{Carlesi} et~al.}{2014}]{Carlesi:2013a}
{Carlesi} E.,  {Knebe} A.,  {Lewis} G.~F.,  {Wales} S.,    {Yepes} G.,  2014,
  \mnras, 439, 2943

\bibitem[\protect\citeauthoryear{{Carlesi}, {Knebe}, {Lewis} \&
  {Yepes}}{{Carlesi} et~al.}{2014}]{Carlesi:2013b}
{Carlesi} E.,  {Knebe} A.,  {Lewis} G.~F.,    {Yepes} G.,  2014, \mnras, 439,
  2958

\bibitem[\protect\citeauthoryear{{Chiba}, {De Felice} \& {Tsujikawa}}{{Chiba}
  et~al.}{2013}]{Chiba:2013}
{Chiba} T.,  {De Felice} A.,    {Tsujikawa} S.,  2013, \prd, 87, 083505

\bibitem[\protect\citeauthoryear{{Clampitt}, {Cai} \& {Li}}{{Clampitt}
  et~al.}{2013}]{Clampitt2013}
{Clampitt} J.,  {Cai} Y.-C.,    {Li} B.,  2013, \mnras, 431, 749

\bibitem[\protect\citeauthoryear{{Dawson} et~al.,}{{Dawson}
  et~al.}{2013}]{Dawson2013}
{Dawson} K.~S.,  et~al., 2013, \aj, 145, 10

\bibitem[\protect\citeauthoryear{{Hamaus}, {Sutter} \& {Wandelt}}{{Hamaus}
  et~al.}{2014}]{Hamaus2014}
{Hamaus} N.,  {Sutter} P.~M.,    {Wandelt} B.~D.,  2014, ArXiv e-prints:
  1403.5499

\bibitem[\protect\citeauthoryear{{Jennings}, {Li} \& {Hu}}{{Jennings}
  et~al.}{2013}]{Jennings2013}
{Jennings} E.,  {Li} Y.,    {Hu} W.,  2013, \mnras, 434, 2167

\bibitem[\protect\citeauthoryear{Kreckel, {Ryan Joung} \& Cen}{Kreckel
  et~al.}{2011}]{Kreckel2011}
Kreckel K.,  {Ryan Joung} M.,    Cen R.,  2011, \apj, 735, 132

\bibitem[\protect\citeauthoryear{{Lavaux} \& {Wandelt}}{{Lavaux} \&
  {Wandelt}}{2010}]{Lavaux2010}
{Lavaux} G.,  {Wandelt} B.~D.,  2010, \mnras, 403, 1392

\bibitem[\protect\citeauthoryear{Li}{Li}{2011}]{Li2011}
Li B.,  2011, \mnras, 441, 2615

\bibitem[\protect\citeauthoryear{{Li}, {Zhao} \& {Koyama}}{{Li}
  et~al.}{2012}]{Li2012}
{Li} B.,  {Zhao} G.-B.,    {Koyama} K.,  2012, \mnras, 421, 3481

\bibitem[\protect\citeauthoryear{Li \& Zhao}{Li \& Zhao}{2009}]{Li2009}
Li B.,  Zhao H.,  2009, \prd, 80

\bibitem[\protect\citeauthoryear{Muller, Arbabi-Bidgoli, Einasto \&
  Tucker}{Muller et~al.}{2000}]{Muller2000}
Muller V.,  Arbabi-Bidgoli S.,  Einasto J.,    Tucker D.,  2000, \mnras, 318,
  280

\bibitem[\protect\citeauthoryear{Neyrinck}{Neyrinck}{2008}]{Neyrinck2008}
Neyrinck M.~C.,  2008, \mnras, 386, 2101

\bibitem[\protect\citeauthoryear{Nusser, Gubser \& Peebles}{Nusser
  et~al.}{2005}]{Nusser:2004}
Nusser A.,  Gubser S.~S.,    Peebles P.,  2005, Phys.Rev., D71, 083505

\bibitem[\protect\citeauthoryear{{Pan}, {Vogeley}, {Hoyle}, {Choi} \&
  {Park}}{{Pan} et~al.}{2012}]{Pan2011}
{Pan} D.~C.,  {Vogeley} M.~S.,  {Hoyle} F.,  {Choi} Y.-Y.,    {Park} C.,  2012,
  \mnras, 421, 926

\bibitem[\protect\citeauthoryear{{Peebles}}{{Peebles}}{2001}]{Peebles2001}
{Peebles} P.~J.~E.,  2001, \apj, 557, 495

\bibitem[\protect\citeauthoryear{{Peebles} \& {Ratra}}{{Peebles} \&
  {Ratra}}{1988}]{Peebles:1988}
{Peebles} P.~J.~E.,  {Ratra} B.,  1988, \apjl, 325, L17

\bibitem[\protect\citeauthoryear{{Pisani}, {Lavaux}, {Sutter} \&
  {Wandelt}}{{Pisani} et~al.}{2013}]{Pisani2013}
{Pisani} A.,  {Lavaux} G.,  {Sutter} P.~M.,    {Wandelt} B.~D.,  2013, ArXiv
  e-prints: 1306.3052

\bibitem[\protect\citeauthoryear{{Planck Collaboration}}{{Planck
  Collaboration}}{2013}]{Planck2013}
{Planck Collaboration} 2013, ArXiv e-prints: 1303.5076

\bibitem[\protect\citeauthoryear{{Ratra} \& {Peebles}}{{Ratra} \&
  {Peebles}}{1988}]{Ratra:1988}
{Ratra} B.,  {Peebles} P.~J.~E.,  1988, \prd, 37, 3406

\bibitem[\protect\citeauthoryear{{Reid} et~al.,}{{Reid}
  et~al.}{2012}]{Reid2012}
{Reid} B.~A.,  et~al., 2012, \mnras, 426, 2719

\bibitem[\protect\citeauthoryear{{Spolyar}, {Sahl{\'e}n} \& {Silk}}{{Spolyar}
  et~al.}{2013}]{Spolyar2013}
{Spolyar} D.,  {Sahl{\'e}n} M.,    {Silk} J.,  2013, ArXiv e-prints: 1304.5239

\bibitem[\protect\citeauthoryear{{Springel}}{{Springel}}{2005}]{Springel:2005}
{Springel} V.,  2005, \mnras, 364, 1105

\bibitem[\protect\citeauthoryear{{Sutter} et~al.,}{{Sutter}
  et~al.}{2014}]{Sutter2014c}
{Sutter} P.~M.,  et~al., 2014, ArXiv e-prints: 1406.1191

\bibitem[\protect\citeauthoryear{{Sutter}, {Lavaux}, {Wandelt} \&
  {Weinberg}}{{Sutter} et~al.}{2012a}]{Sutter2012b}
{Sutter} P.~M.,  {Lavaux} G.,  {Wandelt} B.~D.,    {Weinberg} D.~H.,  2012a,
  \apj, 761, 187

\bibitem[\protect\citeauthoryear{{Sutter}, {Lavaux}, {Wandelt} \&
  {Weinberg}}{{Sutter} et~al.}{2012b}]{Sutter2012a}
{Sutter} P.~M.,  {Lavaux} G.,  {Wandelt} B.~D.,    {Weinberg} D.~H.,  2012b,
  \apj, 761, 44

\bibitem[\protect\citeauthoryear{{Sutter}, {Lavaux}, {Wandelt}, {Weinberg} \&
  {Warren}}{{Sutter} et~al.}{2013}]{Sutter2013c}
{Sutter} P.~M.,  {Lavaux} G.,  {Wandelt} B.~D.,  {Weinberg} D.~H.,    {Warren}
  M.~S.,  2013, ArXiv e-prints: 1310.7155

\bibitem[\protect\citeauthoryear{{Sutter}, {Lavaux}, {Wandelt}, {Weinberg} \&
  {Warren}}{{Sutter} et~al.}{2014}]{Sutter2013b}
{Sutter} P.~M.,  {Lavaux} G.,  {Wandelt} B.~D.,  {Weinberg} D.~H.,    {Warren}
  M.~S.,  2014, \mnras, 438, 3177

\bibitem[\protect\citeauthoryear{{Sutter}, {Lavaux}, {Wandelt} B.~D.~Hamaus,
  {Weinberg} \& {Warren}}{{Sutter} et~al.}{2013}]{Sutter2013a}
{Sutter} P.~M.,  {Lavaux} G.,  {Wandelt} B.~D.~Hamaus N.,  {Weinberg} D.~H.,
  {Warren} M.~S.,  2013, ArXiv e-prints: 1309.5087

\bibitem[\protect\citeauthoryear{{Sutter} \& Ricker}{{Sutter} \&
  Ricker}{2008}]{Sutter2008}
{Sutter} P.~M.,  Ricker P.~M.,  2008, \apj, 687, 7

\bibitem[\protect\citeauthoryear{{Tavasoli}, {Vasei} \& {Mohayaee}}{{Tavasoli}
  et~al.}{2013}]{Tavasoli2013}
{Tavasoli} S.,  {Vasei} K.,    {Mohayaee} R.,  2013, \aap, 553, A15

\bibitem[\protect\citeauthoryear{Tinker \& Conroy}{Tinker \&
  Conroy}{2009}]{Tinker2009}
Tinker J.~L.,  Conroy C.,  2009, \apj, 691, 633

\bibitem[\protect\citeauthoryear{{Wang}, {Chen} \& {Chen}}{{Wang}
  et~al.}{2012}]{Wang:2012}
{Wang} P.-Y.,  {Chen} C.-W.,    {Chen} P.,  2012, \jcap, 2, 16

\end{thebibliography}
}

\end{document}